\documentclass[aps,prd,10pt,twocolumn,superscriptaddress,preprintnumbers]{revtex4}

\usepackage{bm}
\usepackage{braket}

\usepackage{blindtext}
\usepackage{epsfig, cancel}

\usepackage{latexsym}
\usepackage{natbib, comment}
\usepackage{mathrsfs,amsmath,amssymb,amsthm,amsfonts,tikz,graphicx,accents,hyperref,color}
\usepackage{url}
\usepackage{dcolumn}
\usepackage{multirow}
\usepackage{color}
\usepackage{cancel}
\usepackage{soul}
\usepackage[normalem]{ulem}
\usepackage{txfonts}
\usepackage{epsfig}
\usepackage{psfrag}
\usepackage{subfigure}
\hypersetup{colorlinks=true}
\usepackage{mathtools}
\usepackage{enumitem}
\usepackage{float}

\def\H0{{\text{H}\hspace*{-2.05mm}\text{H} 0\hspace*{-1.35mm}0\ }}

\usepackage[all]{xy}
\usepackage{slashed}
\usepackage{slashed,ccaption}
\usepackage{multirow}

\usepackage{caption}

\usepackage{array}
\hypersetup{ linktoc=all,
    colorlinks, linkcolor={palatinateblue},
    citecolor={brightpink}, urlcolor={amaranth}
}

\graphicspath{{Images/}}

\renewcommand{\d}[1]{\ensuremath{\operatorname{d}\!{#1}}}

\DeclareSymbolFont{extraup}{U}{zavm}{m}{n}
\DeclareMathSymbol{\varheart}{\mathalpha}{extraup}{86}
\DeclareMathSymbol{\vardiamond}{\mathalpha}{extraup}{87}
\makeatletter
\renewcommand*{\@fnsymbol}[1]{\ensuremath{\ifcase#1\or \clubsuit \or \vardiamond \or \varheart\or
    \spadesuit\or \mathparagraph\or \|\or **\or \dagger\dagger
    \or \ddagger\ddagger \else\@ctrerr\fi}}
\makeatother

\definecolor{rosy}{RGB}{230,235,252}
\definecolor{myframetitle}{RGB}{90,89,170}
\definecolor{myblocktitle}{RGB}{140,185,249}
\definecolor{mytitle}{RGB}{10,80,26}

\definecolor{darkgreen}{RGB}{27,130,45}
\definecolor{darkblue}{rgb}{0,0,0.3}
\definecolor{darkred}{rgb}{0.7,0,0}

\definecolor{light gray}{RGB}{220,220,220}
\definecolor{dark purple}{RGB}{108,0,217}
\definecolor{pink}{RGB}{190,20,100}
\definecolor{orang}{RGB}{193,63,0}
\definecolor{green}{RGB}{11,98,17}
\definecolor{darkpink}{RGB}{153,0,76}
\definecolor{bluegreen}{RGB}{0,102,102}
\definecolor{greenlagan}{RGB}{0,102,0}
\definecolor{redgreen}{RGB}{102,102,0}
\definecolor{Redgreen}{RGB}{153,76,0}
\definecolor{vividviolet}{rgb}{0.62, 0.0, 1.0}
\definecolor{amaranth}{rgb}{0.9, 0.17, 0.31}
\definecolor{palatinateblue}{rgb}{0.15, 0.23, 0.89}
\definecolor{brightpink}{rgb}{1.0, 0.0, 0.5}
\definecolor{cornflowerblue}{rgb}{0.39, 0.58, 0.93}
\definecolor{deepcarminepink}{rgb}{0.94, 0.19, 0.22}
\definecolor{radicalred}{rgb}{1.0, 0.21, 0.37}

\newcommand{\bTh}{\boldsymbol{\Theta }}

\newcommand{\bO}{\boldsymbol{\Omega}}

\DeclareFontFamily{OT1}{rsfs}{}

\DeclareFontShape{OT1}{rsfs}{m}{n}{ <-7> rsfs5 <7-10> rsfs7 <10->rsfs10}{} 

\DeclareMathAlphabet{\mycal}{OT1}{rsfs}{m}{n}

\newcommand{\be}{\begin{equation}}
\newcommand{\ee}{\end{equation}}

\begin{document}



\title{Gravitational Stress Tensor and Current at Null Infinity in Three Dimensions
}
\author{H.~Adami}\email{hamed.adami@bimsa.cn} 
\affiliation{Shanghai Institute for Mathematics and Interdisciplinary Sciences (SIMIS), Shanghai, 200433, China}
\affiliation{Yau Mathematical Sciences Center, Tsinghua University, Beijing 100084, China}
\affiliation{Beijing Institute of Mathematical Sciences and Applications (BIMSA), Huairou District, Beijing 101408, P. R. China}
\author{M.~M.~Sheikh-Jabbari}\email{jabbari@theory.ipm.ac.ir}
\affiliation{School of Physics, Institute for Research in Fundamental Sciences (IPM), P.O.Box 19395-5531, Tehran, Iran}
\author{V.~Taghiloo}\email{v.taghiloo@iasbs.ac.ir}
\affiliation{School of Physics, Institute for Research in Fundamental
Sciences (IPM), P.O.Box 19395-5531, Tehran, Iran}
\affiliation{Department of Physics, Institute for Advanced Studies in Basic Sciences (IASBS),
P.O. Box 45137-66731, Zanjan, Iran}

\begin{abstract}

We develop the framework that reveals the intrinsic conserved stress tensor and current associated with the null infinity of a  three-dimensional ($3d$) asymptotically flat spacetime. These are, respectively,  canonical conjugates of degenerate metric and  Ehresmann connection of the boundary Carrollian geometry. Their conservation reproduces the Bondi-mass and angular momentum conservation equations if the asymptotic boundary is endowed with a torsional affine connection that we specify. Our analysis and results shed further light on the $3d$ flat holography; the stress tensor and current give rise to an asymptotically flat fluid/gravity correspondence. The requirement of a well-defined $3d$ action principle yields Schwarzian action at null infinity governing the dynamics induced by reparametrizations over the celestial circle, in accord with the codimension $2$ holography of $3d$ flat spacetimes.

\end{abstract}
\maketitle

\section{Introduction}

Defining a gravitational energy-momentum tensor (EMT) has posed a significant challenge within the framework of general relativity, largely due to the subtlety arising from the equivalence principle \cite{Ashtekar:1974uu} (see \cite{Szabados:2009eka} for a review). Among various proposals, the Brown-York quasi-local energy-momentum tensor \cite{brown1993quasilocal} emerges as a particularly promising candidate. This tensor finds interpretation as the EMT of a boundary theory \cite{Balasubramanian:1999re, Henningson:1998gx}. It plays a pivotal role in the holographic description of AdS backgrounds, especially in the fluid/gravity correspondence \cite{Bhattacharyya:2007vjd, Bhattacharyya:2008xc, Haack:2008cp, Rangamani:2009xk}.

Brown-York approach is applicable to time-like or space-like boundaries, e.g. asymptotic boundary of Anti-de Sitter (AdS) spacetimes \cite{Balasubramanian:1999re, Henningson:1998gx, Papadimitriou:2004ap, Papadimitriou:2005ii, Compere:2019bua, Compere:2020lrt, Fiorucci:2020xto, Alessio:2020ioh, Adami:2023fbm}.
On the other hand, null boundaries, such as black hole horizons and light-like boundaries of asymptotically flat spacetimes, are more interesting. Defining the EMT for null boundaries presents a significant challenge due to the degenerate nature of the induced boundary metric.  Recently, there has been notable progress in defining the EMT for null boundaries at finite distances \cite{Parattu:2015gga, Jafari:2019bpw, Chandrasekaran:2021hxc}. These developments have renewed interest in exploring concepts like the membrane paradigm \cite{Price:1986yy, Grumiller:2022qhx} and delving deeper into black hole physics \cite{Grumiller:2018scv, Adami:2021nnf, Adami:2021kvx, Freidel:2022bai, Redondo-Yuste:2022czg, Freidel:2022vjq, Adami:2023fbm, Adami:2023wbe, Odak:2023pga, Ciambelli:2023mir}.

Flat space holography \cite{Bagchi:2012xr, Bagchi:2012yk, Afshar:2013vka, Bagchi:2013lma, Detournay:2014fva} and fluid/gravity correspondence for asymptotically flat spacetimes necessitates defining an EMT on null infinity \cite{Donnay:2022aba, Donnay:2022wvx, Ciambelli:2024kre}. Previous efforts to construct such an EMT have used two different approaches: (1) Constructing null EMT from a large radius limit of AdS \cite{Campoleoni:2022wmf, Ciambelli:2020eba} and resorting to notions like the auxiliary rigging vectors \cite{Chandrasekaran:2021hxc, Freidel:2022vjq, Freidel:2022bai, Freidel:2024tpl}. (2) Considering the Carrollian limit of a theory and study the Carrollian conservation laws \cite{Ciambelli:2018wre, Ciambelli:2018ojf, Ciambelli:2018xat, Donnay:2019jiz}.

In this letter, we determine the gravitational stress tensor and current on null infinity in 3-dimensional ($3d$) asymptotically flat spacetimes, relying exclusively on the full \emph{intrinsic} geometry of the null boundary, without resorting to any limit process. Null infinity is a Carrollian geometry specified by the \emph{Carrollian geometric-triple}, a degenerate metric, a kernel, and a dual vector to the kernel which is usually called the Ehersmann connection. This yields the existence of a stress tensor and a canonical momentum, a null current, that are respectively canonically conjugate to the metric and the dual vector. 
Moreover, covariance under Carrollian boosts fixed the equation-of-state (EoS) of the Carrollian fluid to pressure equal minus the energy density, in accord with results in \cite{deBoer:2021jej, deBoer:2023fnj}.

A key result of our analysis is that the Carrollian structure of the null infinity \cite{Bagchi:2022eav, Bagchi:2022nvj} should be supplemented with a \emph{torsional connection} \cite{Bergshoeff:2017btm, Hansen:2021fxi}. The connection is fixed requiring the Carrollian \emph{geometric-triple compatibility} conditions which involve covariant consistency of the triple as well as invariance of local frame fixing for kernel vector and its dual 1-form. The geometric-triple compatible connection yields an interesting physical result: conservation of the hydrodynamic entities with respect to the torsional connection should yield the Bondi mass and angular momentum continuity equations.  

We also discuss that having a well-defined $3d$ variational principle and the presence of a conserved stress tensor on null infinity, require an additional boundary action \footnote{A similar construction for EMT in $3d$ flat spacetimes with derivative expansions is discussed by Hartong \cite{Hartong:2015usd}. However, due to the more constrained geometric structure, no boundary action emerged in that work.}.  Remarkably, the boundary action manifests as a Schwarzian action on the celestial circle, effectively capturing the dynamics induced by reparametrizations of the celestial circle. This result corroborates the codimension $2$ holography proposed for $3d$ asymptotically flat spacetimes. 

This Letter is organized as follows. We first review the construction of solution phase space for flat $3d$ gravity and geometry of the null boundary of asymptotically flat spacetimes and discuss the hydrodynamic aspect of the symplectic potential. Subsequently, we introduce and utilize a suitable torsional connection to formulate components of Einstein's equation projected at the boundary as conservation laws for a stress tensor and a current. Requiring variational principle, we infer that the boundary action is Schwarzian. Finally, we close by concluding remarks and outlook. In the appendices, we discuss how our analyses and results can be retrieved from an infinite radius limit on the AdS$_3$ background.

\section{Solution phase space}\label{sec:solution-phase-space}
We first review the construction of the solution phase space for pure $3d$ flat space gravity as detailed in \cite{Adami:2022ktn, Adami:2023fbm, Geiller:2021vpg}. Such a theory is described by a Lagrangian of the form 
\begin{equation}
    L[g]:= \frac{1}{16\pi G}\, \sqrt{-g}\,  R +\partial_\mu L_{\mathcal{B}}^{\mu}\, ,
\end{equation}
where $g$ denotes determinant of spacetime metric $g_{\mu\nu}$, $R$ is the Ricci scalar and $L_{\mathcal{B}}^{\mu}[g]$ stands for boundary Lagrangian. Taking first-order variation of the Lagrangian, one can read equations of motion, Ricci tensor equals zero $R_{\mu\nu}=0$, and the symplectic potential \cite{lee:1990nz, Iyer:1994ys, Wald:1999wa}
\begin{equation}\label{Theta-mu-generic}
  \Theta^\mu[ g ; \delta g]:=\Theta_{_{\text{LW}}}^\mu [ g ; \delta g] + \nabla_{\nu} Y^{\mu \nu}[ g ; \delta g]+\delta  L_{\mathcal{B}}^{\mu}[ g ] \, .
\end{equation}
where $\Theta^{\mu}_{_{\text{LW}}} [g; \delta g]$ is the Lee-Wald symplectic potential for pure gravity,  
\begin{equation}\label{Theta-LW}
    \Theta^{\mu}_{_{\text{LW}}} [g; \delta g]:=\frac{\sqrt{-g}}{8 \pi G} \nabla^{[\alpha} \left( g^{\mu ] \beta} \delta g_{\alpha \beta} \right)\, ,
\end{equation}
and $Y^{\mu\nu}$ is a skew-symmetric tensor constructed out of metric and its variation. We refer to this tensor as $Y-$freedom and shall fix it by physical requirements. 

Let us adopt a Gaussian-null-type coordinate system $x^\mu=\{v, r, \phi \}$  where $v$ denotes the advanced time, $r$ the radial coordinate, and $\phi \sim \phi +2\pi$ the polar coordinate. In this coordinate system, line-element is
\begin{equation}\label{metric}
    \d s^2= g_{\mu\nu}\d x^\mu \d x^\nu= -V \d v^2 + 2 \eta \d v \d r + {\cal R}^2 \left( \d \phi + U \d v \right)^2\, ,
\end{equation}
where $\eta=\eta(v,\phi)$  while $V,U,\mathcal{R}$ are generic functions on the spacetime. Three components of Einstein's equations completely specify the $r$ dependence of these functions, \footnote{The differential operator $\mathcal{D}_v$ which acts on a codimension one function $O_w$ of weight $w$ is defined through
$
    \mathcal{D}_v O_w:= \partial_v O_w -\mathcal{U} \partial_\phi O_w - w O_w \partial_\phi \mathcal{U}  \, .
$
For example $\mathcal{M},\Upsilon,\Omega, \Pi$ are of weight $2,2,1,1$ respectively.}
\begin{subequations}\label{metric-components}
    \begin{align}
    \mathcal{R}=& \Omega + r \, \eta \, \lambda \, ,\\
     U= &\ {\cal U} +  \frac{1}{\lambda \, {\cal R}}\, \frac{ \eta'}{ \eta }  + \frac{{8 G \Upsilon}-\Omega\, \Pi'}{2 \lambda {\cal R}^2} \, ,\\
          V= & \ \frac{1}{\lambda^2}\Biggl[  -8 G \mathcal{M}-2\, \text{Sch}[\sigma;\phi]+ {\lambda\, \Omega\, \mathcal{D}_{v}\Pi}+ \left(\frac{\eta'}{\eta}\right)^2 \nonumber\\
          &\hspace{-0.6 cm}+\frac{({8 G \Upsilon}-\Omega\, \Pi')^2}{4  {\cal R}^2} - \frac{2\mathcal{R}}{\eta }\,    \mathcal{D}_v ( \eta \lambda ) +\left(\frac{{8 G \Upsilon}-\Omega\, \Pi'}{\mathcal{R}}\right) \frac{ \eta'}{ \eta } \Biggr] \, ,
    \end{align}
\end{subequations}
where prime denotes derivative with respect to $\phi$ and 
\begin{subequations}\begin{align}
\Pi & :=2 \ln | \eta \lambda\Omega^{-1} |\\
    \text{Sch}[\sigma;\phi] & :=\frac{ \sigma'''}{\sigma'}-\frac{3}{2}\left(\frac{\sigma''}{\sigma'}\right)^2\, ,\qquad \sigma:= \int^\phi \lambda \d \phi\, ,
\end{align}
\end{subequations}
where $\text{Sch}[\sigma;\phi]$ is the Schwarzian derivative. There are six  functions of $v,\phi$ in \eqref{metric-components}, namely $\{ \eta,  \mathcal{U}, \Omega, \lambda, \mathcal{M}, \Upsilon \}$, where $\mathcal{M}$ and $\Upsilon$ are respectively Bondi mass and angular momentum aspects. The two remaining components of Einstein’s equations impose two constraints among these functions
\begin{subequations}\label{M-Upsilon-EoM}
\begin{align}
    &{\mathcal{D}_{v}\mathcal{M}+\frac{1}{4G}\, \mathcal{U}'''=0}\, , \label{EOM-M}\\
    &{\mathcal{D}_{v}\Upsilon-\lambda \left(\frac{ \mathcal{M}}{\lambda^2}\right)'+\frac{1}{4G}\,(\lambda^{-1})'''=0} \, .\label{EOM-Upsilon}
\end{align}
\end{subequations}
These equations are first order equations in boundary time $v$ and may hence be viewed as ``constraint equations''. Moreover, they involve third order derivative terms in $\phi$, which we call ``anomalous terms''. The solution space is hence specified by 4 functions of $v,\phi$, that may conveniently be chosen ${\cal M}, \Upsilon, \Pi, \Omega$; \eqref{M-Upsilon-EoM} may be solved for $\mathcal{U}, \lambda$. 

\noindent\textbf{Symmetry generators.} The vector field 
\begin{equation}\label{null-bondary-sym-gen}
    \xi = T\, \lambda\, \partial_{v}+\left[Y-\mathcal{U}\, \lambda \, T+\frac{(T\, \lambda)'}{\lambda\, \mathcal{R}}\right]\partial_{\phi} + \xi^r \partial_r\, ,
\end{equation}
with
\begin{equation}
    \begin{split}
        \xi^r = & \frac{1}{\eta \, \lambda} \left[ Z -T \, \lambda \, \mathcal{D}_v \Omega-(\Omega Y)' - \frac{1}{\eta }\left(\frac{\eta\,(T\lambda)'}{\lambda}\right)'\right]\\
        &- \frac{r }{2}\, \left(  W +  T\, \lambda \,  \mathcal{D}_v \Pi -2\, e^{\Pi/2} Z +Y\, \, \Pi'\right)\\
        &-\frac{{8 G \Upsilon}-\Omega \, \Pi' }{2\, \eta\, \lambda^2\, \mathcal{R}} \, (T\lambda)'  \, ,
    \end{split}
\end{equation}
moves us within the solutions space constructed above and hence is the symmetry generator in the usual sense of the Noether theorem. 
This vector field is parametrized by 4 functions of $v,\phi$: supertranslations $T(v,\phi)$ and $Z(v,\phi)$, superscaling $W(v,\phi)$, and superrotation $Y(v,\phi)$. 

Analyses in \cite{Adami:2022ktn, Adami:2023fbm}  revealed that ${\cal M}, \Upsilon, \Pi, \Omega$ are charge aspects respectively associated with $T, Y, Z, W$ symmetry generators with the transformation laws,
\begin{subequations}\label{delta-fields}
    \begin{align}
    {\delta_{\xi} \mathcal{M} \approx} &\ {Y\,  \mathcal{M}'+2\,  \mathcal{M}\,Y'-\frac{1}{4G}\, Y'''\, , }\\
    {\delta_{\xi} \Upsilon \approx} &\ {T\,  \mathcal{M}'+2\,  \mathcal{M}\, T'+Y\, \Upsilon'+2\,  \Upsilon\, Y'-\frac{1}{4G}\, T'''\, ,}\\
      \delta_{\xi} \Omega = &\  Z\, , \qquad
   \delta_{\xi}\Pi=\ - W\, ,
   \end{align}
\end{subequations} 
where $\approx$ indicates on-shell equality in which equations of motion \eqref{M-Upsilon-EoM} are used.

\noindent\textbf{Symplectic form and charges.} 
To discuss the symplectic form, we should fix $Y$-freedom and boundary Lagrangian $ L_{\mathcal{B}}^\mu$.  We fix the former upon the requirement that symplectic potential and hence the surface charge variations are finite and $r$ independent. This is achieved through
\begin{equation}\label{Y-r-removing}
    Y^{\mu\nu}[\delta g; g]=\frac{1}{8 \pi G}\left(2\delta \sqrt{-g}\, n^{[\mu}l^{\nu]}+3\sqrt{-g}\, \delta n^{[\mu}l^{\nu]} \right)\, ,
\end{equation}
where $l^\mu$ and $n^\nu$ are two null vector fields
\begin{equation}\label{null-basis}
    l^{\mu}\partial_{\mu} =\partial_{v}+\frac{V}{2\eta}\partial_{r}-U\partial_{\phi}\, , \qquad n^{\mu}\partial_{\mu}=-\frac{1}{\eta}\partial_{r}\, ,
\end{equation}
such that they are normalized as $n \cdot l =-1$. The boundary Lagrangian will be fixed later, requiring a well-defined variational principle. With this $Y$-term, the symplectic potential on arbitrary constant $r$ surfaces takes the form 
\begin{equation}\label{Symp-Pot}
\bTh:=  \int \Theta^\mu \d{}^2 x_\mu= \bTh_{\mathcal{H}} +\bTh_{\mathcal{C}}+(\textit{total variation terms})
\end{equation}
with 
\begin{subequations}\label{symplectic-pot-total}
    \begin{align}
         \bTh_{\mathcal{H}}:= &\   -{\frac{1}{2 \pi }} \int \d{}^2 x \left[  \mathcal{M} \, \delta (\lambda^{-1}) +  \Upsilon \, \delta \mathcal{U} \right]\, , \label{symplectic-pot-hydro}\\
         \bTh_{\mathcal{C}}:= &\  \frac{1}{16 \pi G} \int  \d{}^2 x \, \partial_v\left( \Omega \, \delta \Pi \right)\, , 
    \end{align}
\end{subequations}
where $\bTh_{\mathcal{H}}$ is the ``hydrodynamical part'' and $\bTh_{\mathcal{C}}$ is the ``corner part'', is a total derivative in $v$ and may be computed at a constant $v=v_b$ section on the presumed constant $r$ boundary. The contributions from the boundary Lagrangian $ L_{\mathcal{B}}^{\mu}$ appear in the ``total variation term''.

We employ the covariant phase space formalism \cite{lee:1990nz, Iyer:1994ys, Wald:1999wa, Grumiller:2022qhx} to read the symplectic form and surface charges:
\begin{equation}\label{surface-charge-Y-infty}
\begin{split}
&\bO [\delta g ,\delta g; g] :=\delta \bTh[\delta g; g]\, ,\\
&\delta Q_\xi :=\bO [\delta g ,\delta_\xi g; g]  = \frac{1}{16\pi G} \int_0^{2\pi} \d \phi \left( W\, \delta\Omega +Z \, \delta \Pi\right)\\
&\qquad\qquad +\frac{1}{2\pi} \int_0^{2\pi} \d \phi \left( T\, \delta  \mathcal{M}+Y \, \delta  \Upsilon   \right)\, .
\end{split}
\end{equation}
where $\delta$ denotes the exterior derivative on the solution phase space. 
The charge algebra is the direct sum of the $\mathfrak{Heisenberg}$ and the $\mathfrak{bms}_3$ algebras \cite{Adami:2022ktn, Adami:2023fbm}. The former is spanned by $\Omega$ and $\Pi$ while the latter is spanned by $ \mathcal{M}$ and $ \Upsilon$.

\section{Carrollian geometry of null boundary}\label{sec:CGoNB}
So far we assumed a boundary at a constant but arbitrary $r$. We now examine the geometric structure of null infinity,  setting the stage for the analysis in subsequent sections. To this end, we consider the induced metric at null infinity $\mathcal{I}$, that may be defined as $r\to \infty$ limit of spacetime metric divided  by $r^2$, up to an arbitrary conformal factor \cite{Ashtekar:1978zz, Ashtekar:2014zsa}
\begin{equation}
    \d s^2|_{_{\mathcal{I}}} := \left(\frac{\omega}{\eta\, \lambda}\right)^2 \lim_{r\to \infty} \frac{\d s^2}{r^2}:= \gamma_{a b} \d x^a \d x^b\, ,
\end{equation}
where
\begin{equation}
    \gamma_{a b}=k_{a} k_{b}\d x^a  \d x^b,\qquad k_{a} \d x^a:=\omega \left( \d \phi + \mathcal{U} \d v \right) \, ,
\end{equation}
and $\omega=\omega(v,\phi)$ is an arbitrary function at null boundary.  In what follows we choose $\omega=1$. 

The induced metric defined above is degenerate, i.e. there exists a ``kernel vector field'' $l^a$  such that $\gamma_{ab} l^b=0$ (or $l^a k_a=0$) and its explicit form is given by 
\begin{equation}\label{l-def}
    l^{a}\partial_{a}:= \alpha^{-1}(\partial_{v}-\mathcal{U}\partial_{\phi})\, ,
\end{equation} 
where $\alpha(v,\phi)$ is an arbitrary function (at null boundary). The Ehresmann connection may be defined as
\begin{equation}\label{n-def}
n_{a}\d{}x^a=-\alpha \d{}v+b \,  k_a \d x^a\, , \qquad l^{a}n_{a}=-1\, ,
\end{equation} 
where $b(v,\phi)$ is an arbitrary function and stands for a Carrollian boost,  we choose $b=0$. $l^a, n_a$ defined above are related to null vectors  $l^\mu,n_\nu$ in \eqref{null-basis}  by a pullback and a rescaling, namely $\alpha^{-1} l^{a} \to {l}^{a}$ and $\alpha\,n_{a} \to  {n}_{a}$. Induced metric $\gamma_{ab}$, kernel $l^a$, and Ehresmann connection $n_a$ form a \emph{ruled} Carrollian structure, constituting the \textit{Carrollian geometric-triple} $(\gamma_{ab}; l^a, n_a)$, \emph{cf.} \cite{Duval:2014uoa, Duval:2014uva, Duval:2014lpa, Henneaux:1979vn, Henneaux:2021yzg, Ciambelli:2019lap,deBoer:2021jej, deBoer:2017ing} for more discussions.

The projection map
\begin{equation}\label{Proj-def}
P^{a}{}_{b}:=\delta_{b}^{a}+n_{b}l^{a}\, , \qquad P^{a}{}_{b}l^{b}= P^{a}{}_{b}n_{a}=0\, ,
\end{equation} 
allows us to define partially inverse metric $h^{ab}$, such that $h^{ac}\gamma_{cb}=P^{a}{}_{b}$. Note that there is a freedom in the definition of $h^{ab}$ that can be fixed through the condition $h^{ab}n_b=0$, yielding
\begin{equation}\label{h-def}
h^{ab}=k^{a}k^{b}\, , \qquad k^{a}\partial_{a}:=\partial_{\phi}\, .
\end{equation}
Note that the vector $k^a$ and one-form $k_a$ are related as $k^{a}=h^{ab}k_{b}$ and they satisfy $k^ak_a=1, k^a n_a=0$.

Finally, the volume form on null infinity may be defined through the minors of $\gamma_{ab}$ defined as $\mathfrak{g}^{ab}= \varepsilon^{ac}\varepsilon^{be}\gamma_{ce}$, where $\varepsilon^{ab}$ is the $2d$ Levi-Civita symbol and $\mathfrak{g}^{ab}\gamma_{bc}=0$. The minors can be expressed as $\mathfrak{g}^{ab}=\gamma \, l^a l^b$ where $\gamma=\alpha^2$ plays the role of determinant of metric when the metric is degenerate \cite{Henneaux:1979vn, Henneaux:2021yzg}.

\section{Hydrodynamic symplectic potential}
Given the Carrollian geometric-triple discussed in the previous section, especially since the geometry at null infinity is characterized by a degenerate metric $\gamma_{ab}$ and a co-vector $n_a$, we revisit the hydrodynamic aspect of the symplectic potential \eqref{symplectic-pot-hydro} and rewrite it in terms of a stress tensor and a current. Motivated by surface charges in \eqref{surface-charge-Y-infty}, we define a $(1,1)$-type stress-tensor and a null current as
\begin{equation}\label{non-anomalous-EMT}
\begin{split}
    \mathcal{T}^{a}{}_{b}&=-\mathcal{P}\, k^{a}k_{b}-\Upsilon \, l^{a}k_{b}\, , \qquad 
    p^{a}=\mathcal{M}\, l^{a}\, ,\\
\mathcal{M}:=-n_{a}p^{a},&\qquad \Upsilon:=n_a\mathcal{T}^{a}{}_{b}k^b,\qquad \mathcal{P}:=-k_a\mathcal{T}^{a}{}_{b}k^b
\end{split}
\end{equation}
where {$\mathcal{M}, \Upsilon, \mathcal{P}$} respectively denote  the energy/mass aspect, the angular momentum aspect, and the pressure. One may also define the $(2,0)$-type stress tensor
\begin{equation}
    \mathcal{T}^{ab}:=h^{bc} \mathcal{T}^{a}{}_{c}= -\mathcal{P}\,  k^a k^b -{\Upsilon}\,  l^a k^b \, .
\end{equation}
Once we choose $\alpha=\lambda^{-1}$, the hydrodynamic symplectic potential can be written as
\begin{equation}\label{hydro-sym-pot-EMT}
    \bTh_{\mathcal{H}}={\frac{1}{2\pi}}\int \d{}^2 x \sqrt{\gamma}\left( \mathcal{T}^{ab}  \delta \gamma_{ab}+ p^{a}  \delta n_{a}\right)\, .
\end{equation}

Recalling that $k^a \delta k_a = 0$ and in the chosen gauges for $\omega, \alpha$ and $b$ (the conformal and  Carroll boost frames), the pressure $\mathcal{P}$ remains undetermined. It can be fixed by requiring  an additional condition such as 
\begin{equation}\label{EoS}
    \mathcal{T}^a{}_a+p^{a}n_{a}=0\, \quad \Longrightarrow\quad {\mathcal{M}+\mathcal{P}=0}\, . 
\end{equation}  
Here ${\mathcal{P}=-\mathcal{M}}$ is the EoS for Carrollian fluid \cite{deBoer:2021jej,deBoer:2023fnj} and is consistent with the conservation of stress tensor, which we will discuss later. To understand this EoS, we recall the notion of the ``tilt'' in a fluid, the momentum flow of the fluid on adapted constant time slices \cite{King:1972td}. Our $p^a$ may also be viewed as a tilt. Fluids with EoS  $\mathcal{M}+\mathcal{P}=0$ are not affected by the tilt \cite{Krishnan:2022qbv}. This fact guarantees the (Carrollian) boost invariance of the null boundary fluid. As a closing remark, we note that  $\mathcal{T}^a{}_b$ tensor and $p^a$ vector in the null boundary case may be related to the symmetric (Brown-York) EMT of the timelike boundary of AdS$_3$ case, which has a non-degenerate metric. This has been discussed in the appendix.

\section{Geometric-triple compatible connection}
To explore conservation laws we need to supplement the ruled Carrollian geometric-triple by defining a connection on the null boundary. Consider a generic affine connection $\Gamma^c_{ab}$ and define covariant derivative of a generic tensor $\mathcal{X}^{a\cdots}_{b \cdots}$ as
\begin{equation}
    D_c \mathcal{X}^{a\cdots}_{b \cdots}=\partial_c \mathcal{X}^{a\cdots}_{b \cdots} + \Gamma^a_{cd} \mathcal{X}^{d\cdots}_{b \cdots}+\cdots- \Gamma^d_{cb} \mathcal{X}^{a\cdots}_{d \cdots}-\cdots \, .
\end{equation}

While connection is in general a geometric information independent of the metric, it is usually fixed by metric compatibility condition. In the Carrollian case, geometric information is encoded in the triple  $(\gamma_{ab}, l^a; n_a)$, or equivalently in  $(\gamma_{ab}, l^a; P^a{}_b)$. If $a,b=1,\cdots, d$ this triple may be viewed as the information needed to embed a $d$ dimensional null surface in a $d+1$ dimensional ambient spacetime. As discussed in the previous section, $l^a, n_a$ may be viewed as a frame vector and a 1-form (two of $d+1$ vielbein of the embedding space). In contrast to the usual geometries, the triple contains information about the metric and a frame 1-form. The frame vector and 1-form in our triple are fixed in a certain local conformal frame (by choice of $\omega=1, b=0, \alpha=1/\lambda$). Recalling that (spin) connection is not a tensor under local frame transformations, our ``Carrollian geometric-triple compatibility'' conditions, besides covariant constancy of the triple, should also contain conditions to guarantee the local frame fixing. The latter conditions should specify the Lie derivative of the Carrollian geometry triples along the kernel vector $l^a$, denoted by $\mathcal{L}_{l}$. The  Carrollian geometric-triple compatibility conditions may be expressed as
\begin{subequations}\label{Carrollian-metric-compatibility-conditions}
\begin{align}
D_c \gamma_{ab}=0\, , \quad & \quad P^d{}_c  D_{d}  P^{a}{}_{b} =0\, ,\qquad { P^b{}_c D_b l^a =0} \, , \label{CMCC-a}\\
& K_{ab}:= \frac{1}{2}\mathcal{L}_{l}\gamma_{ab} ={ \frac{1}{2}}D_c l^c\  \gamma_{ab} \, , \label{CMCC-b} \\ 
&\mathcal{L}_l P^a{}_b= \frac{1}{2} l^c D_c P^a{}_{b}\, ,\label{CMCC-c}
\end{align}
\end{subequations}
{where \eqref{CMCC-b} and \eqref{CMCC-c} ensure our choice of local conformal Carrollian frame.}

The above conditions may be solved to obtain
\begin{equation}\label{connection}
    \Gamma^{c}_{ab}= \frac{1}{2}h^{cd}\left(\partial_{a}\gamma_{bd}+\partial_{b}\gamma_{ad}-\partial_{d}\gamma_{ab}\right)+h^{cd}\,K_{da}n_{b}+l^c S_{ab}\, ,
\end{equation}
where $h^{ab}$ is the partially inverse metric and 
\begin{equation}\label{S-ab}
    S_{ab}:=-3\,\partial_{(a}n_{b)}{- n_{[a} \mathcal{L}_l n_{b]}}+2\partial_{c}l^{c}\, n_{a}n_{b}\, .
\end{equation}

Note that conditions \eqref{Carrollian-metric-compatibility-conditions} can be used to define a generic Carrollian connection, not specific to $2d$ cases ($3d$ gravity). Below we make some comments for the case of our main focus in this work, $3d$ gravity, where $n_a, l^a, k_a$ may be used as frame basis: \begin{enumerate}
    \item $D_c \gamma_{ab}=0$  yields $D_a k_b=0$. This condition fixes the first two terms in  \eqref{connection}; this does not fix the $S_{ab}$ term.
\item $D_c \gamma_{ab}=0, P^d{}_c D_{d} P^{a}{}_{b} =0$ condition yields $P^a{}_b D_{a} k^c=0=P^d{}_c D_{d} h^{ab}$ and we have
\begin{equation}
    D_a k^b = -\theta_k \, n_a \, l^b,\qquad \theta_k:= D_a k^a
\end{equation}
\item $P^a{}_b D_{a} l^c=0$  and \eqref{connection} yield
\begin{equation}
    D_a l^b = -\theta_l \, n_a \, l^b,\qquad \theta_l:= D_a l^a
\end{equation}
\item Eq.~\eqref{Carrollian-metric-compatibility-conditions} yield $P^a{}_b D_{a} n_c=0$. Moreover, we have, 
\begin{equation}
    D_a n_b = n_a (\theta_l \,  n_b - \theta_k \,  k_b)\, .
\end{equation}
\item $S_{ab}=-\Gamma^{c}_{ab} n_c$ has symmetric and antisymmetric parts and its antisymmetric part is proportional to $k_{[a} n_{b]}$. 
\item The connection \eqref{connection} is torsional with the torsion tensor,
\begin{equation}\label{Torsion}
\begin{split}
    T^c_{ab}:= 2\Gamma^c_{[ab]}=& 2 h^{cd}\, K_{d[a}n_{b]}+2 l^c S_{[ab]}\, \\
=&    (\theta_l k^{c}+ \theta_k  l^c) k_{[a}n_{b]}\, .
\end{split}
\end{equation}
\end{enumerate}

While \eqref{Carrollian-metric-compatibility-conditions} are geometric requirements, they have physical consequences: divergence of the null current and the stress tensor \eqref{non-anomalous-EMT}, $D_a p^a$, and $ D_{a}\mathcal{T}^{a}{}_{b}$, respectively reproduce non-anomalous parts of constraint equations \eqref{EOM-M} and \eqref{EOM-Upsilon}, i.e. 
\begin{subequations}\label{non-anomalous-parts}
    \begin{align}
        {  \sqrt{\gamma}\,   D_{a}p^{a}} &{= \mathcal{D}_{v} \mathcal{M}}\, , \\
      -  \sqrt{\gamma}\,  D_{b}\mathcal{T}^{b}{}_{a} & =\left[ \mathcal{D}_{v} \Upsilon-\lambda\, \left(\frac{ \mathcal{M}}{\lambda^2}\right)' \right] k_a\, .
    \end{align}
\end{subequations}

\section{Conservation equations}
Supplementing the Carrollian geometric-triple with an appropriate connection \eqref{connection}, we are ready to explore conservation equations of hydrodynamic quantities. Expansions of vectors $l^a, k^a$, namely $\theta_l, \theta_k$, for our solutions are
\begin{equation}\label{expansions}
    \theta_l=-2\, \lambda \,  \mathcal{U}' \, , \qquad \theta_k ={-2\,  \lambda^{-1} \lambda'} \, .
\end{equation}
The immediate distinction between the Carrollian structure we present here and those in earlier literature \cite{Hartong:2015xda, Hartong:2015usd, Bergshoeff:2017btm, Hansen:2021fxi, Bekaert:2015xua, deBoer:2023fnj}
is that we relax conditions $D_a l^a =0\, , D_a k^a=0$, i.e. we relax torsion-free condition (\textit{cf.} \eqref{Torsion}). {Note that $S_{ab}$ reduces to the one in classical Carroll geometry when $\theta_l=\theta_k=0$, \emph{c.f.} \cite{deBoer:2023fnj}.}
As mentioned earlier, \eqref{non-anomalous-parts} only capture non-anomalous parts in \eqref{M-Upsilon-EoM}. 

To write down \eqref{M-Upsilon-EoM} as conservation equations we need to account for the anomalous parts. To this end, we add  the ``anomaly'' parts $\mathcal{A}^{a}{}_{b}, A^{a}$,
\begin{equation}\label{cons-EMT-total}
    \begin{split}
       \text{T}^{a}{}_{b}&:= \mathcal{T}^a{}_b+\mathcal{A}^a{}_b= \mathcal{T}^a{}_b+ \mathcal{A}\,  k^a k_b\\
       \text{P}^{a}&:=p^{a}+A^{a} = ({\cal M}+ \mathcal{A}) \, l^a+ \mathcal{B} \, k^a
    \end{split}
\end{equation}
with
\begin{subequations}
    \begin{align}
     \mathcal{A}:=\mathcal{A}^a{}_{a}  = &\ -\frac{1}{8 G}\left( k^c \partial_c \theta_{k}+\frac{1}{4}\theta_{k}^{2}\right)=\frac{1}{4 G}\mathrm{Sch}[\sigma;\phi]\, , \label{A-B}\\
       \mathcal{B}:=k_a A^a{} =&\ -{\frac{1}{8 G}}\left(k^c \partial_c \theta_l-l^c \partial_c \theta_k\right)\, ,
    \end{align}
\end{subequations}
where in the  last equality in \eqref{A-B} we used \eqref{expansions}. 
The conservation equations 
\begin{equation}\label{totla-T-P-conservation}
    D_{b}\text{T}^{b}{}_{a}=0\, , \qquad D_{a}P^{a}=0\, , 
\end{equation}
reproduce \eqref{M-Upsilon-EoM}.  Conservation equations for \eqref{cons-EMT-total} are the flat counterpart of the EMT conservation equations for the AdS case discussed in \cite{Adami:2023fbm, Campoleoni:2018ltl, Ciambelli:2020ftk, Campoleoni:2022wmf} and may be obtained as the flat limit of the AdS case, \textit{cf}. the appendices. Observe that 
\begin{equation}\label{T-P-EoS}
\begin{split}
 \text{T}^a{}_a&=\mathcal{T}^a{}_a+\mathcal{A}=-{\cal P}+\frac{1}{4 G}\mathrm{Sch}[\sigma;\phi]\, , \\ P^{a} n_{a}&=p^a n_a -\mathcal{A}=-({\cal M}+\frac{1}{4 G}\mathrm{Sch}[\sigma;\phi])\, , 
\end{split}
\end{equation}
therefore, recalling \eqref{EoS},  $\text{T}^a{}_a+P^{a} n_{a}= \mathcal{T}^a{}_a+p^a n_a={\cal M}+{\cal P}$. That is, the equation of state ${\cal P}=-{\cal M}$ is not affected by the anomalous parts $\text{T}^a{}_b, P^a$.

Formulating the intrinsic stress tensor and current \eqref{cons-EMT-total} at null infinity, we have accomplished our primary goal in this letter. Their conservation with respect to the intrinsic \textit{torsional} null boundary connection \eqref{connection} results in the continuity equations for Bondi mass and angular momentum \eqref{M-Upsilon-EoM}. We emphasize that the null nature (Carrollian structure) of the boundary necessitates the presence of a stress tensor and a null vector, rather than a single EMT. In the following, we establish that these are canonically conjugate to the degenerate metric on null infinity $\gamma_{ab}$ and Ehresmann connection $n_a$.

\section{Boundary Schwarzian action}

The boundary Lagrangian $L_{\mathcal{B}}$, while absent in the symplectic form and hence in the surface charges computed in the covariant phase space formalism, appears in the symplectic potential. Explicitly, the total symplectic potential \eqref{Symp-Pot} in terms of $\text{T}^{ab}= \text{T}^a{}_ch^{cb}$ and $\text{P}^{a}$ takes the form 
\begin{equation}\label{SP-01}
        \begin{split}
            \bTh &= {\frac{1}{2\pi}}\int \d{}^2 x \sqrt{\gamma} \left( \text{T}^{ab} \delta \gamma_{ab}+ P^{a} \delta n_{a}\right)+\bTh_{\mathcal{C}} \\
            &+ \delta \int \d{}^2 x \left(L_{\mathcal{B}}+ \frac{\mathcal{M}}{2\pi \sigma'}+\frac{\mathrm{Sch}[\sigma;\phi]}{8\pi G \sigma'} \right)\, ,
        \end{split}
\end{equation}
where $L_{\mathcal{B}}=L_{\mathcal{B}}^\mu s_\mu$ with $s_\mu$ unit normal to constant $r$ surfaces.

Variational principle, the requirement that variation of the action including the boundary terms should vanish on-shell, implies vanishing of total variation term in the above, thus fixing $L_{\mathcal{B}}$:
\begin{equation}\label{Boundry-Lagrangian}
    S_{\mathcal{B}}
    =\ - \int \d v \int_0^{2\pi} \frac{\d \phi}{2\pi \sigma'} \left(  \mathcal{M}+\frac{1}{4 G}\, \mathrm{Sch}[\sigma;\phi]\right)\, .
\end{equation}
Thus, a Schwarzian action emerges as the boundary action. This action effectively captures the $\phi$-reparametrization over the celestial circle.  While ${\cal M}, \sigma$ are functions of $v,\phi$, the action only involves $\phi$-derivatives, in accord with the codimension 2 holographic description of the asymptotically flat spacetimes (for comprehensive reviews on this subject see \cite{strominger2018lectures, Pasterski:2021rjz, McLoughlin:2022ljp, Raclariu:2021zjz, Pasterski:2021raf, Prema:2021sjp}).

Boundary action \eqref{Boundry-Lagrangian} consists of two terms;  ${\cal M}$ encodes bulk dynamics and the Schwarzian derivative term, the boundary dynamics. Following the usual results \cite{Gibbons:1976ue},  boundary action \eqref{Boundry-Lagrangian} may be written in a ``thermodynamical form'',  
\begin{equation}
\begin{split}
  S_{\mathcal{B}}&= -\int \d{}^2 x\ \beta \, \mathcal{G}\, ,\\
  \mathcal{G}:=\mathcal{M}&+(4 G)^{-1}\, \mathrm{Sch}[\sigma;\phi]\, , \qquad  \beta:=(2\pi \sigma')^{-1}\, ,
\end{split}
\end{equation}
where ${\cal G}$ is Gibbs free energy density and $\beta$ denotes inverse of Frolov-Thorne temperature \cite{Frolov:1989jh}. It is instructive to note that as \eqref{T-P-EoS} shows, ${\cal G}=\text{T}^a{}_a=-P^a n_a$, i.e. Gibbs free energy is the trace of the boundary total stress tensor, or equivalently the component of momentum flow along time-direction. Upon a coordinate transformation $\phi\to \sigma$, i.e. taking $\sigma$ to be the coordinate spanning the circle, and recalling that  $\mathrm{Sch}[\sigma;\phi]=-(\sigma')^2 \mathrm{Sch}[\phi;\sigma]$ with $\sigma=\sigma(\phi)$, the Frolov-Thorne temperature becomes $(2\pi)^{-1}$ and boundary Lagrangian \eqref{Boundry-Lagrangian} becomes a pure Schwarzian derivative  (without $\sigma'$ in the denominator) \cite{Joung:2023doq, Lee:2024etc}.

It is worth noting that the Schwarzian action also appears as the boundary action in the context of $2d$ JT gravity  \cite{Maldacena:2016upp}. A similar discussion on the stretched horizon has appeared in \cite{Carlip:2022fwh}. In this case, the boundary is $1d$ and it describes time-reparametrization over a cut-off on the hyperbolic plane, whereas in our case the Schwarzian action involves derivatives in $\phi$. The emergence of the Schwarzian action at null infinity of asymptotically flat spacetimes is anticipated due to the Carrollian nature of the asymptotic null boundary that has a degenerate and effectively $1d$  metric. Furthermore, it is the simplest term that remains invariant under the global SL$(2,\mathbb{R})$ part of superrotation transformations.

Finally, the exterior derivative of symplectic potential \eqref{SP-01} on the solution phase space yields the symplectic form 
\begin{equation}
        \begin{split}
            \bO =&\ {\frac{1}{2\pi}}\int \d{}^2 x  \left[ \delta (\sqrt{\gamma}\, \text{T}^{ab}) \wedge \delta \gamma_{ab}+ \delta (\sqrt{\gamma} P^{a}) \wedge \delta n_{a}\right] \\
            &+\frac{1}{16 \pi G} \int \d{}^2x\, \partial_v \left(  \delta \Omega \wedge \delta \Pi \right)\, .
        \end{split}
\end{equation}
The first line denotes the hydrodynamic (codimension 1) and the second line the codimension 2 part (corner term) of symplectic potential \cite{Adami:2023wbe, Sheikh-Jabbari:2022mqi}.

\begin{center}
\textbf{Gibbons-Hawking-York term at null boundary}
\end{center}
We have shown that the requirement of a well-defined variational principle yields the boundary Schwarzian action.  On the other hand, the Gibbons-Hawking-York (GHY) boundary term is usually employed to assert the variational principle with the Dirichlet boundary condition. This boundary term is usually written for time or space-like boundaries. The GHY term for null boundaries \cite{Parattu:2015gga, Chandrasekaran:2021hxc, Jafari:2019bpw, Oliveri:2019gvm, Aghapour:2018icu} is given by
\begin{equation}
    S_{{\text{\tiny{GHY}}}}=\frac{1}{8 \pi G}\int \d{}^{2}x\, \sqrt{q_{\phi\phi}}\, \Theta_{l}\, ,
\end{equation}
The null expansion is defined as $\Theta_l :=q_{\mu\nu}\nabla^{\mu} l^\nu$ where $q_{\mu\nu}=g_{\mu\nu}+l_\mu n_\nu+l_\nu n_\mu$ is induced metric on co-dimension 2 surface, $l^{\mu}$ is given in  \eqref{null-basis} and $\sqrt{q_{\phi\phi}}=\mathcal{R}$ is the volume form on a null surface at a finite constant $r$ hypersurface.

To explore the boundary action at null infinity, we may start from $S_{{\text{\tiny{GHY}}}}$ and take the large $r$ limit. Therefore, (for more details see \cite{Adami:2023fbm})
\begin{equation}
\begin{split}
    \Theta_{l}=-\frac{1}{2r{\eta}\lambda^2}\big(8G\, {\mathcal{M}} &+2\mathcal{S}[\tilde{\sigma};\phi]\big)+\mathcal{O}(r^{-2})\,, \\
     \sqrt{q_{\phi\phi}}&=\eta\lambda\, r+ {\cal O}(1) \, .
\end{split}
\end{equation}
The null GHY action is hence 
\begin{equation}\label{BA-01}
    S_{{\text{\tiny{GHY}}}}=-\int \d{}^{2}x\, \frac{\eta}{2\pi \tilde{\sigma}'}\left(\, {\mathcal{M}}+(4G)^{-1}\mathcal{S}[\tilde{\sigma};\phi]\right)\, ,
\end{equation}
here we defined $\tilde{\sigma}':=\eta\lambda$. {This boundary action matches with our result in \eqref{Boundry-Lagrangian} when we write it in the conformal frame introduced in section \ref{sec:CGoNB}. Explicitly, the choice of conformal frame is equivalent to fixing $\eta=1$ in \eqref{BA-01} and hence $\tilde{\sigma}\to \sigma$.}

\section{Outlook}\label{sec:discussion}

We reviewed the construction of a general solution space for $3d$ Ricci flat spacetimes which involves 4 functions over the null boundary \cite{Adami:2022ktn}. This is a specific subset of the most general case discussed in \cite{Grumiller:2017sjh}. This solution space may be obtained as the flat limit of a similar solution space in  AdS$_3$ \cite{Adami:2023fbm, Geiller:2021vpg, Grumiller:2016pqb}.

We focused on describing the physics of the solution space from the boundary observer's viewpoint. Since the boundary of $3d$ flat space is a null surface, it is described by a Carrollian geometry. We completed the usual ruled Carrollian geometry specified by the triple $(\gamma_{ab}, l^a; P^a{}_b)$, or equivalently $(\gamma_{ab}, l^a; n_a)$, by constructing a ``geometric-triple compatible'' connection, requiring conditions \eqref{Carrollian-metric-compatibility-conditions}. These yield a torsional connection with the torsion being proportional to the expansion of kernel vector $l^a$ and the other frame vector $k_a$,  respectively $\theta_l, \theta_k$. As we have discussed in the appendix, this Carrollian geometry may be obtained as a flat space limit of AdS$_3$. 

Equipped with these geometric notions, we constructed an ``intrinsic'' conserved stress tensor and current at null infinity that are canonical conjugates to the Carrollian degenerate metric $\gamma_{ab}$ and the Ehresmann connection $n_a$. We discussed the hydrodynamical meaning of the solution space through the total stress tensor and the current, given in \eqref{cons-EMT-total}. Conservation of these quantities, which is a manifestation of the Bondi mass and angular momentum conservation equations is a consequence of geometric-triple compatible connection (the connection satisfying \eqref{Carrollian-metric-compatibility-conditions}).  

Some comments on total stress tensor and current \eqref{cons-EMT-total} are in order:
(1) This Carrollian fluid has EoS  pressure+energy=0, a manifestation of the Carrollian boost invariance of the fluid. (2) The (2,0)-type stress tensor is not symmetric, a trait it shares with Carrollian EMTs \cite{deBoer:2021jej}. 
(3) Unlike the AdS$_{3}$ case, which exhibits the trace anomaly \cite{Henningson:1998gx, Adami:2023fbm}, the flat case lacks any trace anomaly. This characteristic bears resemblance to logarithmic celestial conformal field theories (CFTs with zero central charge) \cite{Fiorucci:2023lpb}. (4) In the literature there exists alternative ``non-intrinsic'' approaches to the same problem: (I) Starting with the Brown-York EMT at a finite radius in asymptotically flat spacetimes and taking the radius to infinity \cite{Ciambelli:2018wre, Ciambelli:2018ojf, Ciambelli:2018xat, Donnay:2019jiz}; (II) Starting with the Brown-York EMT for AdS$_{3}$ boundaries and then taking the flat space limit of AdS$_{3}$ \cite{Adami:2023fbm, Campoleoni:2018ltl, Ciambelli:2020ftk, Campoleoni:2022wmf, Ciambelli:2020eba, Chandrasekaran:2021hxc, Freidel:2022vjq, Freidel:2022bai, Freidel:2024tpl}. A comparative analysis of these methodologies with our intrinsic analysis and results is postponed to future work.

In our Carrollian geometric-triple description $(\gamma_{ab}, l^a; n_a)$ and the geometric-triple compatible connection defined through \eqref{Carrollian-metric-compatibility-conditions}, we fixed the ``local'' conformal Carrollian symmetry (counterpart of local Lorentz transformation in the usual $3d$ gravity) by fixing the scaling $\alpha$, the boost $b$ and the $2d$ Weyl scaling $\omega$. This was done through our choices for the triads $l^a, n_a, k^a$. It is interesting to explore how the addition of these three symmetry generators enters in our symplectic form and charge analysis. The first steps in this direction have been taken in \cite{Grumiller:2017sjh, Geiller:2020okp, Prabhu:2021cgk}. 

The requirement of a well-defined variational principle fixed the boundary Lagrangian to be given by a Schwarzian action. We also showed that the same boundary action may be obtained from the Gibbons-Hawking-York term at the asymptotic null boundary. {This boundary Lagrangian is manifestly invariant under time-dependent superrotations $\phi\to \phi+Y(v,\phi)$ while is anomalous under time-dependent supertranslations $v\to v+ T(v,\phi)$. We expect that Weyl scaling $\omega$ and local gauge transformations, $\alpha$ and $b$, can compensate for the anomaly \cite{Odak:2022ndm}. Hence, it is plausible that one could reinstate BMS$_3$ invariance.}
Recalling that $\phi$ is a periodic direction, this theory has a thermal description; it is equal to Gibbs free energy density times the inverse of the Frolov-Thorne temperature. Explicitly, our solution space is described by 4 functions on the null boundary, two of which (${\cal M}, \Upsilon$) specify stress tensor and current and the other two ($\Omega, \Pi$) describe boundary degrees of freedom that are a thermal system at Frolov-Thorne temperature,  governed by the Schwarzian action.

Besides the asymptotic null boundary, black hole horizons are also null boundaries of spacetimes as viewed by non-inertial observers outside the black hole, e.g. see \cite{Adami:2020ugu, Adami:2021nnf}.  Geometrically, the asymptotic and the horizon null boundaries are respectively related working with outgoing and infalling Gaussian null coordinates which may respectively be viewed as Carrollian and infinite boost limits of the usual $2d$ CFT. In $3d$ case these two are very similar \cite{Andringa:2010it, Bagchi:2012xr, FarahmandParsa:2018ojt, Bergshoeff:2020fiz, Bagchi:2022owq, Grumiller:2023rzn} but in general they are very different.  The former is more like a In\"on\"u-Wigner contraction \cite{inonu1953contraction} (speed of light to infinity) and the latter like a Carrollian (speed of light to zero) limit of a usual Lorentzian theory. In $3d$ case both of the limits yield the same BMS$_3$ algebra, which is obtained as a contraction of the Virasoro $\oplus$ Virasoro algebra. One should, however, note that the Witt subalgebra of the BMS$_3$ in these two cases correspond to two different ``diagonal'' subalgebra of the two Virasoro's. \footnote{Near horizon limit of extremal black holes, in particular extremal $3d$ BTZ black holes, has been proposed to be dual to a chiral CFT \cite{Guica:2008mu}. This chiral CFT corresponds to a different contraction of the Virasoro $\oplus$ Virasoro algebra, associated with the Discrete Light Cone Quantization (DLCQ) of the $2d$ CFT \cite{Balasubramanian:2009bg}.} Moreover, states and Hilbert spaces of the null boundary and the horizon theories are associated with different coadjoint orbits/representations of the BMS$_3$ algebras. See for more discussions \cite{Oblak:2016eij} and references therein. These points of course deserve further understanding.

Finally, and most importantly, it is desirable to study how the ``intrinsic'' geometric approach of this work for formulation of the gravitational stress tensor on null infinity, and hence a fluid/gravity correspondence for asymptotically flat spacetimes, extends to the higher dimensional cases, in particular to $4d$.

\begin{acknowledgments}
We would like to thank Luis Apolo, 
Kedar S. Kolekar, Ali Parvizi, Mohammad Hassan Vahidinia,
and Hossein Yavartanoo for discussions or comments on the manuscript.
The work of HA is supported
by Beijing Natural Science Foundation under Grant No IS23018 and by the National Natural Science Foundation of China under Grant No 12150410311. The work of MMShJ is in part supported by the INSF grant No 4026712.
\end{acknowledgments}

\appendix

\section{From AdS to flat}
Here we discussed (1) the Carrollian geometric-triple; (2) constructed the triple compatible connection; (3) worked out the stress tensor and the current and studied its conservation. 
As a consistency check for our analyses and results, we explore if these three aspects can be reproduced from a flat space limit of the AdS$_3$ results.

To start, we review the results for asymptotically AdS$_3$ spacetimes \cite{Adami:2023fbm}. The line element on the conformal boundary of  asymptotically AdS$_3$ spacetimes is given by
\begin{equation}\label{rel:metric-asy}
    \d {\hat{s}}^{2}=\hat{\gamma}_{ab}\d x^{a}\d x^{b}={-\frac{1}{\ell^2\, \lambda^2}  \d v^2}+(\d\phi+\mathcal{U}\d v)^2\, .
\end{equation}
We note that $\sqrt{-\hat{\gamma}}=(\ell\, \lambda)^{-1}$.
Metric at infinity can be expressed in terms of zwibein,
\begin{subequations}
\begin{align}
&\hat{\gamma}_{ab}  = -  \hat{t}_{a} \hat{t}_{b}+\hat{k}_{a} \hat{k}_{b}\, , \\ 
&\hat{t}^{a}\partial_{a}
       =\ell \, \lambda \, (\partial_{v}-\mathcal{U}\partial_{\phi})\, , \qquad   \qquad \hat{t}_{a}\d x^a =- \frac{1}{\ell \lambda} \d v \, , \\
&   \hat{k}^{a}\partial_{a}
   =\partial_{\phi}\, , \qquad  \qquad \hat{k}_{a}\d x^a =   \d \phi+\mathcal{U} \d v\, , 
\end{align}
\end{subequations}
with $\hat{t}^2= -1,\ \hat{k}^2=1,\ \hat{t}\cdot \hat{k}=0$. 

The deviation tensors associated with these vectors are
\begin{subequations}\label{hat-theta}
    \begin{align}
        & \hat{\nabla}_a \hat{t}_{b}=-\hat{\theta}_{k}\, \hat{t}_{a}\hat{k}_{b}+\hat{\theta}_{t}\, \hat{k}_{a}\hat{k}_{b}\, , \quad \hat{\nabla}_a \hat{k}_{b}=- \hat{\theta}_{k}\, \hat{t}_{a}\hat{t}_{b}+ \hat{\theta}_{t}\, \hat{k}_{a}\hat{t}_{b} \, ,\\
    &\hat{\theta}_{k}:=\hat{\nabla}_{a}\hat{k}^{a}=-\lambda^{-1}\lambda'\, , \quad \hat{\theta}_{t}:=\hat{\nabla}_{a}\hat{t}^{a}=-\ell\, \lambda\, \mathcal{U}'\, ,
    \end{align}
\end{subequations}
where $\hat{\nabla}_{a}$ is the covariant derivative with respect to boundary metric $\hat{\gamma}_{ab}$.
The total symmetric, divergence-free energy-momentum tensor for the asymptotically AdS$_{3}$ spacetimes with the boundary metric \eqref{rel:metric-asy} is given by 
\begin{equation}\label{T-to-T'} 
\hat{\text{T}}^{ab}:=\hat{\mathcal{T}}^{ab}-\hat{\mathcal{A}}^{ab}, \quad \hat{\text{T}}^a{}_{a}=-\frac{c}{12}\, \hat{R}\, ,\quad \hat\nabla_b\hat{\text{T}}^{ab}=0\, .
\end{equation}
where 
\begin{subequations}
    \begin{align}
        & \hat{\mathcal{T}}^{ab}:=- \ell\mathcal{M}\, \hat{t}^a \hat{t}^b + \Upsilon\hat{k}^{a} \hat{t}^{b }+\Upsilon\hat{k}^{b} \hat{t}^{a}-\ell\mathcal{M}\, \hat{k}^{a} \hat{k}^{b} \\
        & \hat{\mathcal{A}}^{ab}:=-\frac{c}{12 }\, \Bigl[\hat{\theta}_{t}^2 \, ( \hat{t}^{a}\hat{t}^{b}+\hat{k}^{a}\hat{k}^{b})-\hat{R} \, \hat{k}^{a}\hat{k}^{b} \nonumber\\
        & \qquad \qquad \qquad +2\,  \hat{k}^c\partial_c\hat{\theta}_{t}\,( \hat{t}^a\hat{k}^{b} + \hat{t}^b\hat{k}^a)\Bigr]\,, \\
        & \hat{R}=2 \left( \hat{t}^c  \partial_c \hat{\theta}_{t} + \hat{\theta}_{t}^2 -\hat{k}^c  \partial_c \hat{\theta}_{k} -\hat{\theta}_{k}^2 \right) \, ,
    \end{align}
\end{subequations}
and $c=3\ell/(2G)$ is the Brown-Henneaux central charge \cite{Brown:1986nw}.

\subsection{Carrollian geometric-triple}
The Carrollian geometric-triple $(\gamma_{ab}; l^a, n_a)$ may be obtained through $\ell\to\infty$ limit with the following rescaled quantities:
\begin{equation}\label{re-scale}
         \hat{t}^a \rightarrow  \ell l^a \, , \quad \hat{t}_a \rightarrow  \ell^{-1} n_a \, , \quad \hat{k}^a \rightarrow  k^a \, , \quad \hat{k}_a \rightarrow  k_a \, , 
\end{equation}
where $\gamma_{ab}=k_a k_b$. For later use, we note that, 
\begin{equation}\label{re-scale'}
            \hat{\theta}_{t}\rightarrow \frac{\ell}{2} \theta_l\, , \qquad \hat{\theta}_{k}\rightarrow\frac{1}{2} \theta_k\, , \qquad \sqrt{-\hat{\gamma}}\rightarrow \ell^{-1} \sqrt{\gamma}\, .
\end{equation}
Of course to verify the above one needs the information about the connection which we discuss next.

\subsection{Geometric-triple compatible connection}
Using the above re-definitions, we can rewrite the connection compatible with $\hat{\gamma}_{ab}$ as
\begin{equation}\label{connection-l}
    \begin{split}
        \hat{\Gamma}^c_{ab} 
        &=\  \frac{1}{2}h^{cd}\left(\partial_{a}\gamma_{bd}+\partial_{b}\gamma_{ad}-\partial_{d}\gamma_{ab}\right) \\
        &-l^{c} \Bigl[   \partial_{(a} n_{b)} +n_{(a} \mathcal{L}_l n_{b)}\Bigr] +\ell^2 \, l^{c}  K_{ab}+\frac{\ell^{-2}}{2} \theta_k\,  k^{c}    n_a n_{b}\, .
    \end{split}
\end{equation}
One can also simply check that
\begin{equation}
    \mathcal{L}_{l}\gamma_{ab}= \theta_l \gamma_{ab} \, , \qquad \partial_{[a} n_{b]}=\frac{1}{2}\, \theta_k \, k_{[a} n_{b]}\, .
\end{equation}
Under rescaling, \eqref{hat-theta} become
\begin{equation}\label{D-D}
    \begin{split}
        &\ell^{-1} D_a n_{b} +\ell^{-1}\left(\Gamma^c_{ab}-\hat{\Gamma}^c_{ab}\right)n_c=-\frac{1}{2}\ell^{-1}\theta_{k}\, n_{a}k_{b}+\frac{\ell}{2} \theta_{l}\, k_{a}k_{b}\, ,  \\
        & D_a k_{b} +\left(\Gamma^c_{ab}-\hat{\Gamma}^c_{ab}\right)k_c=- \frac{\ell^{-2}}{2} \theta_{k}\, n_{a}n_{b}+ \frac{1}{2}\theta_{l}\, k_{a}n_{b}\, , \\
        & \ell\,  D_a l^{b}-\ell \left(\Gamma^b_{ac}-\hat{\Gamma}^b_{ac}\right) l^c=-\frac{\ell^{-1}}{2} \theta_{k}\, n_{a}k^{b}+\frac{\ell}{2}\theta_{l}\, k_{a}k^{b}\, , \\
        & D_a k^{b}-\left(\Gamma^b_{ac}-\hat{\Gamma}^b_{ac}\right) k^c=- \frac{1}{2} \theta_{k}\, n_{a} l^{b}+\frac{\ell^2}{2} \,  \theta_{l}\, k_{a}l^{b}\, ,
    \end{split}
\end{equation}
where $D_a$ denotes covariant derivative compatible with $\gamma_{ab}$ which implies that $D_a k_b =0$.
We supplement metric compatibility with local conformal frame-fixing conditions,
\begin{subequations}
    \begin{align}
        & D_a l^b =  -\theta_l\, l^b\,  n_a\, ,   \\
        & D_a n_b = \theta_l \, n_a \, n_b - \theta_k \, n_a \,  k_b\, ,  \\
        & D_a k^b = -\theta_k \, n_a \, l^b\, , \\
        & D_a k_b =0\, .
    \end{align}
\end{subequations}

Plugging the above and \eqref{connection-l} in \eqref{D-D}, we find that
\begin{subequations}
    \begin{align}
    &\hspace{-0.3 cm}\Gamma^c_{ab}k_c= \left(\frac{1}{2}h^{cd}\left(\partial_{a}\gamma_{bd}+\partial_{b}\gamma_{ad}-\partial_{d}\gamma_{ab}\right)+  h^{cd}K_{da } n_{b} \right)k_{c}, \\
     & \hspace{-0.3 cm}  -\Gamma^c_{ab}n_c  =\theta_l \, n_a \, n_b  -\theta_{k}\, n_{a}k_{b} -  \partial_{(a} n_{b)} -n_{(a} \mathcal{L}_l n_{b)}. 
    \end{align}
\end{subequations}
Now we can read the null connection as
\begin{equation}\label{}
    \begin{split}
       & \hspace{-0.3 cm} \Gamma^{c}_{ab}=  P^c{}_{d}\Gamma^{d}_{ab} - l^c n_d \Gamma^{d}_{ab} \\
       & \hspace{-0.1 cm}  = \frac{1}{2}h^{cd}\left(\partial_{a}\gamma_{bd}+\partial_{b}\gamma_{ad}-\partial_{d}\gamma_{ab}\right)+h^{cd}\,K_{da}n_{b}+l^c S_{ab}\, ,
    \end{split}
\end{equation}
where
\begin{equation}
    S_{ab}= -3\,\partial_{(a}n_{b)}{- n_{[a} \mathcal{L}_l n_{b]}}+2\partial_{c}l^{c}\, n_{a}n_{b}\, .
\end{equation}
\subsection{Relation between EMT and stress tensor}
The symplectic potential for asymptotically AdS$_3$ spacetimes then takes the form \cite{Adami:2023fbm}
\begin{equation}
    \begin{split}
\bTh&=-\frac{1}{4\pi} \int \d{}^2 x\ \sqrt{-\hat{\gamma}}\, \hat{\text{T}}^{ab}\, \delta \hat{\gamma}_{a b}\\
&+\frac{1}{16\pi G} \, \int \d{}^2 x\ \partial_v(\Omega\,\delta\Pi) \\
&{+}\int\delta \Biggl[ L_{\mathcal{B}} +\frac{\ell}{{16}\pi G }\, \sqrt{-\hat{\gamma}}\, \hat{\theta}_{t}^2  -\frac{\mathcal{R}^2}{{4}\pi G \ell^2 \lambda}+\frac{\mathcal{M}}{2\pi \lambda}\Biggr]\, .
    \end{split}
\end{equation}
Let us rewrite the first term as 
\begin{equation}
        \begin{split}
            \sqrt{-\hat{\gamma}}\, \hat{\text{T}}^{ab}\, \delta \hat{\gamma}_{a b}  =& -2\sqrt{\gamma}\, \left(\ell^{-2}\hat{\text{T}}^{ab} \hat{t}_b \right)\, \delta n_{a} \\
            &+\sqrt{\gamma}\, \left(\ell^{-1}\hat{\text{T}}^{ab} \right)\, \delta \gamma_{a b}\, .
        \end{split} 
\end{equation}
Under rescaling \eqref{re-scale} and \eqref{re-scale'} we have
\begin{equation}
   \hspace{-0.6 cm} \ell^{-2}\hat{\text{T}}^{ab}\hat{t}_b = \mathcal{M}\, l^a -\frac{1}{8G }  (k^c\partial_c\theta_{l})\, k^a    -\frac{\ell^2}{32 G}\, \theta_{l}^2 \, l^{a} - \ell^{-2} \Upsilon \, k^{a} \,,
\end{equation}
\begin{equation}
\begin{split}
     \hspace{-0.6 cm}   \ell^{-1}\hat{\text{T}}^{ab} & = - \ell^{2} \mathcal{M}\, l^a l^b + \Upsilon k^{a} l^{b }+ \Upsilon k^{b} l^{a}-\mathcal{M}\, k^{a} k^{b} \\
        &+ \frac{\ell^2}{8G }\Bigl[\frac{1}{4}\theta_{l}^2 \, ( \ell^2 l^{a}l^{b}+k^{a}k^{b})-\ell^{-2}\hat{R} \, k^{a} k^{b} \\
        &\qquad\   +  k^c\partial_c \theta_{l}\, (l^a k^{b} +l^b k^a)\Bigr]\, , 
    \end{split}
\end{equation}
or
\begin{subequations}
    \begin{align}
        (\ell^{-2}\hat{\text{T}}^{ab}\hat{t}_b)\delta n_a =& \mathcal{M}\, l^a \delta n_a -\frac{\ell^2}{32 G}\, \theta_{l}^2 \, l^{a} \delta n_a \,, \\
        \frac{1}{2}(\ell^{-1}\hat{\text{T}}^{ab})\delta \gamma_{ab}=&   \Upsilon k^{a} l^{b }\delta \gamma_{ab}
        +\frac{\ell^2}{8 G }\left(k^c\partial_c \theta_{l}
        \right) l^a k^{b} \delta \gamma_{ab} \, .
    \end{align}
\end{subequations}
Using the variation of the boundary term
\begin{equation}
    \begin{split}
        \delta\left(\frac{\ell}{8\pi G }\, \sqrt{-\hat{\gamma}}\hat{\theta}_{t}^2\right)= & \frac{\ell^2}{32\pi G }\,  \delta\left(\sqrt{\gamma}\theta_{l}^2\right) \\
        =& \frac{\ell^2}{32\pi G } \, \sqrt{\gamma}(\theta_l^2 l^a\delta{n_a}+ 4 k^c\partial_c\theta_l l^a\delta k_a) \\
        & -\frac{\ell^2}{8\pi G }\, \partial_\phi(\sqrt{\gamma} \theta_l l^a\delta k_a)\, ,    
    \end{split}
\end{equation}
divergent terms in symplectic potential $\Theta$ cancel out and the finite parts yield the non-anomalous parts of the current and stress tensor. 
Recall that we have already shown how the Carrollian covariant derivative and torsional connection can be obtained in the flat limit and moreover, in the main text we have found the anomalous parts that guarantee the conservation of stress tensor and momentum flow, the anomalous parts are also expected to arise as the flat space limit.

\bibliography{references}

\end{document}